\documentclass[twocolumn,trackchanges]{aastex701}

\begin{document}

\title{Compact Core, Extended Reach: A Bipolar kpc-Scale Elongation in a Little Red Dot at $z \approx 5.5$}

\correspondingauthor{Zhiyuan Ji}
\email{zhiyuanji@arizona.edu}

\author[0000-0001-7673-2257]{Zhiyuan Ji}
\affiliation{Steward Observatory, University of Arizona, 933 N. Cherry Avenue, Tucson, AZ 85721, USA}
\email{zhiyuanji@arizona.edu}

\author[0000-0001-6561-9443]{Yang Sun}
\affiliation{Steward Observatory, University of Arizona, 933 N. Cherry Avenue, Tucson, AZ 85721, USA}
\email{sunyang@arizona.edu}

\author[0000-0002-7831-8751]{Mauro Giavalisco}
\affiliation{University of Massachusetts Amherst, 710 North Pleasant Street, Amherst, MA 01003-9305, USA}
\email{mauro@umass.edu}

\author[0000-0003-3307-7525]{Yongda Zhu}
\affiliation{Steward Observatory, University of Arizona, 933 N. Cherry Avenue, Tucson, AZ 85721, USA}
\email{yongdaz@arizona.edu}

\author[0000-0003-2303-6519]{George H. Rieke}
\affiliation{Steward Observatory, University of Arizona, 933 N. Cherry Avenue, Tucson, AZ 85721, USA}
\email{ghrieke@gmail.com}

\author[0000-0003-2919-7495]{Christina C. Williams}
\affiliation{NSF–DOE Vera C. Rubin Observatory/NSF NOIRLab, 950 N. Cherry Ave., Tucson, AZ 85719, USA}
\affiliation{Steward Observatory, University of Arizona, 933 N. Cherry Avenue, Tucson, AZ 85721, USA}
\email{christina.williams@noirlab.edu}

\author[0000-0003-0695-4414]{Michael V.\ Maseda}
\affiliation{Department of Astronomy, University of Wisconsin-Madison, 475 N. Charter St., Madison, WI 53706 USA}
\email{maseda@wisc.edu}

\author[0000-0002-6221-1829]{Jianwei Lyu}
\affiliation{Steward Observatory, University of Arizona, 933 N. Cherry Avenue, Tucson, AZ 85721, USA}
\email{jianwei@arizona.edu}

\author[0000-0002-7893-6170]{Marcia Rieke}
\affiliation{Steward Observatory, University of Arizona, 933 N. Cherry Avenue, Tucson, AZ 85721, USA}
\email{mrieke@gmail.com}

\author[0000-0002-8224-4505]{Sandro Tacchella}
\affiliation{Kavli Institute for Cosmology, University of Cambridge, Madingley Road, Cambridge, CB3 0HA, UK}
\affiliation{Cavendish Laboratory, University of Cambridge, 19 JJ Thomson Avenue, Cambridge, CB3 0HE, UK}
\email{st578@cam.ac.uk}

\begin{abstract}

Little Red Dots (LRDs) appear extremely compact at rest-frame optical wavelengths, yet many show extended rest-frame UV morphology revealing more complex internal structure.  We present a combined analysis of VLT/MUSE rest-frame UV integral-field spectroscopy and continuum-subtracted [\mbox{O\,{\sc iii}}], H$\beta$, and
H$\alpha$+[\mbox{N\,{\sc ii}}] emission-line maps from JWST/NIRCam imaging at sub-kpc resolution for LRD-204851 at $z=5.482$ in
GOODS-S.  We find that LRD-204851 hosts a remarkably thin, bipolar, elongated structure passing through the optical continuum centroid and extending several kpc on either side, traced by both the UV continuum and the rest-frame optical emission lines, with a bright [\mbox{O\,{\sc iii}}] clump-like structure $\sim$2\,kpc to the south-east of the centroid.  The MUSE observations reveal a double-peaked Ly$\alpha$ profile, with a broad and bright near-systemic red peak and a relatively faint peak blueshifted by $\sim$430\,km\,s$^{-1}$, accompanied by a tentative \mbox{N\,{\sc v}}\,$\lambda 1238$ detection at similar velocity. In narrow-band imaging extracted from the MUSE IFU cube, both the blue Ly$\alpha$ peak and the tentative \mbox{N\,{\sc v}} emission lean toward this same south-eastern direction.  Independently, radiative-transfer modeling of the integrated Ly$\alpha$ profile favors a biconical low-column-density cavity in a dense, slowly expanding neutral envelope, in support of the bipolar geometry traced by the line maps. Together, these results suggest that the elongated emission of LRD-204851 is connected to radiation and/or gas flow from its central engine through a low-column-density channel with a small opening angle that may trace either a slow outflow or a quasi-static ionization cone. LRD-204851 is one of the first LRDs where the central engine's impact on its host galaxy is potentially directly observable on kpc scales.

\end{abstract}



\section{Introduction}
\label{sec:intro}

Little Red Dots (LRDs) are a population of compact, optically red sources discovered in JWST observations \citep{Labbe2023,Barro2024,Kocevski2025,Matthee2024, Greene2024,Furtak2024,Perezgonzalez2024,Williams2024,Akins2025}. They are characterized by a distinctive ``V-shaped'' continuum, blue
in the rest-frame UV and red in the rest-frame optical \citep{Labbe2023,Perezgonzalez2024,Wang2025,deGraaff2025lrdsample}, and a substantial fraction exhibit broad (FWHM~$\gtrsim 1000$~km\,s$^{-1}$) H$\alpha$ emission, often
accompanied by Balmer absorption features \citep{Kokorev2023,
Matthee2024,Maiolino2024,Greene2024,Furtak2024,Juodzbalis2024,
Juodzbalis2026,cliff,Wang2025,JiX2025,Kocevski2025,deGraaff2025lrdsample}.

Despite a rapidly growing observational record, the physical nature of LRDs remains debated.  The broad H$\alpha$ emission has been interpreted as classical Type-1 AGN broad line region (BLR) emission viewed through a partially obscuring neutral medium \citep[e.g.,][]{Matthee2024,Greene2024,Maiolino2024,Juodzbalis2026}, but recent work has also proposed that the broadening is dominated by electron scattering in a dense ionized cocoon surrounding a young supermassive black hole, a ``quasi-star''- or ``black-hole-star''-like configuration \citep[e.g.,][]{Inayoshi2025,cliff,Naidu2025,deGraaff2025lrdsample}. Empirical evidence in support of this picture includes exponential Balmer line wings, ubiquitous Balmer absorbers, and bright broad H$\alpha$ accompanied by H$\beta$ profiles dominated by their narrow component, all of which are difficult to reproduce with classical Keplerian BLR broadening \citep{Rusakov2026nature, Matthee2026}. The nature of the V-shaped continuum is similarly debated, with proposed origins ranging from a dust-reddened AGN \citep{Greene2024,Killi2024,Wang2025} to a super-Eddington intermediate-mass black hole \citep[e.g.,][]{Pacucci2024,Liu2025} to a dust-obscured starburst \citep{Perezgonzalez2024,Williams2024,Akins2025}.  Distinguishing among these scenarios requires diagnostics that probe the kinematics, geometry, and ionization state of the gas around the central source.

Emission lines are a particularly powerful tool for this task, and the rest-frame UV and optical lines probe these properties in complementary ways. Ly$\alpha$ is resonantly scattered, so its emergent profile is especially sensitive to the neutral hydrogen through which it escapes. The offsets and relative strengths of its peaks, together with the shape of its red wing, constrain the \ion{H}{1} column density, kinematics, and geometry  \citep[e.g.,][]{Verhamme2006,Verhamme2008,Hansen2006, Dijkstra2006,lya2026}, quantities that are not directly encoded in the non-resonant optical lines. N\,{\sc v}\,$\lambda1238,1242$ provides a complementary probe of the highly ionized gas. Producing N$^{4+}$ requires photons with energies $\sim77.5$~eV, making N\,{\sc v} particularly sensitive to a hard ionizing spectrum \citep[e.g.,][]{Hainline2011,treiber2025}. Its velocity offset further traces the kinematics of this highly ionized component \citep[e.g.,][]{VandenBerk2001}. In contrast, the rest-frame optical Balmer and forbidden lines, including [\ion{O}{3}]\,$\lambda4959,5007$ ([\ion{O}{3}] hereafter) and [\ion{N}{2}]\,$\lambda6548,6584$ ([\ion{N}{2}] hereafter), are much less affected by resonant scattering. Their relative strengths provide more direct constraints on the excitation and ionization state of the ionized gas, while the Balmer decrement constrains its dust attenuation \citep[e.g.,][]{Baldwin1981,Veilleux1987,Kewley2019}. Combining these diagnostics on the same source therefore connects the neutral gas that regulates the escape of Ly$\alpha$, the kinematics of the highly ionized gas, and the physical conditions of the surrounding ionized medium. Together, they provide a more complete picture of the gaseous environment than any of these diagnostics alone.

Beyond the integrated line measurements, the spatial distribution of emission lines provides a critical, complementary constraint, distinguishing scenarios in which the line emission is co-spatial with the central LRD continuum from those in which it extends to circumnuclear or galactic scales. Recent JWST/NIRSpec IFU observations have demonstrated the importance of such spatially resolved measurements, revealing extended ionized gas around LRDs and providing constraints on its resolved kinematics \citep[e.g.,][]{DEugenio2026a}. JWST/NIRCam imaging is particularly well suited for this purpose at high redshift as well: the rest-frame optical lines \mbox{[\ion{O}{3}]}, H$\beta$, H$\alpha$, and \mbox{[\ion{N}{2}]} redshift into the NIRCam wavelength range.  The growing number of programs with NIRCam medium-band imaging coverage, such as JEMS \citep{Williams2023}, JOF \citep{Eisenstein2025}, MegaScience \citep{Suess2024} and MINERVA \citep{Muzzin2025}, now provide deep coverage in dozens of broad and medium bands across legacy extragalactic fields.  This opens a unique opportunity to construct sub-kpc-resolution maps of the optical emission lines around individual LRDs, complementing ancillary ground-based integral-field rest-frame UV spectroscopy from MUSE \citep[e.g.,][]{Bacon2023} and Keck \citep[e.g.,][]{Erb2023}. 

\begin{figure*}
    \includegraphics[width=0.97\textwidth]{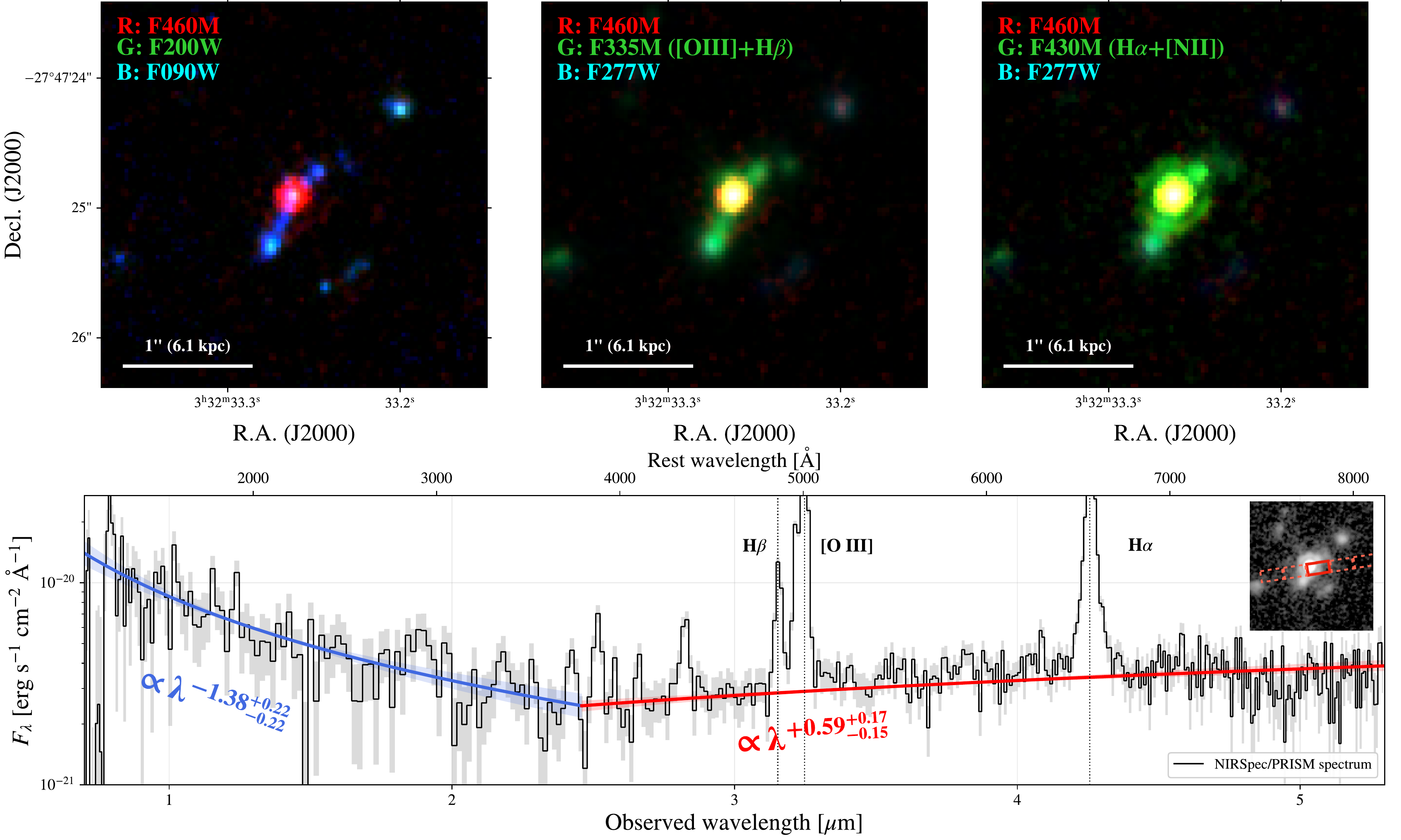}
    \caption{NIRCam imaging and NIRSpec/PRISM spectrum of LRD-204851 at $z_{\rm sys}=5.482$. \textbf{Top row:} Three NIRCam RGB cutouts at the native resolution, all shown with a common linear stretch. The left panel uses three continuum filters; the middle panel highlights [O\,{\sc iii}]+H$\beta$; and the right panel highlights H$\alpha$+[N\,{\sc ii}] in green. The target is a compact red point source in F460M, but shows a remarkably elongated blue continuum extension from southeast to northwest, as well as complex structures in optical emission lines. \textbf{Bottom:} NIRSpec/PRISM spectrum (MSA shuttle is shown in the inset). The blue and red solid lines show the rest-UV and rest-optical continuum power laws, respectively. The characteristic ``V-shaped'' continuum identifies LRD-204851 as a Little Red Dot, and higher-resolution spectra confirm that its H$\alpha$ emission includes a broad component with ${\rm FWHM}>1000~{\rm km~s^{-1}}$ \citep{Matthee2024,Juodzbalis2026,Matthee2026}; these spectra are not shown here.}
    \label{fig:target_info}
\end{figure*}

LRD-204851 (J033233.26-274724.89) is one such source where this combined analysis can be carried out in unprecedented depth.  Spectroscopically confirmed at $z_{\rm sys}=5.482$ based on the \mbox{[\ion{O}{3}]} emission line from NIRSpec high-resolution grating observations \citep{Matthee2026}, the source was first identified as an LRD by \citet{Matthee2024} based on a broad H$\alpha$ emission line detected in a NIRCam/WFSS grism spectrum from the FRESCO survey \citep{oesch2023}.  The broad H$\alpha$ line was subsequently confirmed at higher resolution by NIRSpec R1000 grating observations from the JADES program \citep{Juodzbalis2026}, and the source was further classified as an LRD by \citet{deGraaff2025lrdsample} on the basis of its V-shaped UV-to-optical continuum and compact F444W morphology (also see Figure~\ref{fig:target_info}). Recent NIRSpec/G395H observations further reveal that the broad H$\alpha$ profile is well described by exponential wings with FWHM $\approx$ 1080~km s$^{-1}$ and a blueshifted Balmer absorber at $\Delta v \approx -80$~km s$^{-1}$ \citep{Matthee2026}.

Located close to the Hubble Ultra Deep Field \citep[HUDF;][]{beckwith2006} in GOODS-S \citep{giavalisco2004}, LRD-204851 benefits from one of the most comprehensive multiwavelength datasets available for an LRD, spanning deep JWST NIRCam imaging in 14 broad and medium bands from JADES \citep{Eisenstein2025} and JEMS \citep{Williams2023}, as well as VLT/MUSE spectroscopy from the MUSE Hubble Ultra Deep Field surveys \citep{Bacon2023}.  As Figure~\ref{fig:target_info} shows, while LRD-204851 appears as a compact red point source in the rest-frame optical continuum, the rest-frame UV continuum reveals a remarkably elongated extension from southeast to northwest, and the rest-frame optical emission lines display complex morphologies, features that hint at a rich and structured gas environment around this LRD.

In this Letter, we present a combined analysis of MUSE UV spectroscopy and optical emission-line maps, constructed from JWST/NIRCam imaging, of LRD-204851. We characterize the gas geometry, kinematics, and ionization state around this source. The combination of integrated and spatially resolved UV resonance-line spectroscopy with rest-frame optical emission-line imaging on a single LRD offers a unique multi-wavelength view that sheds light on the competing physical scenarios outlined above, and offers a direct glimpse of a possible central-engine impact on the host galaxy at kpc scales. Beyond this individual source, LRD-204851 provides a pilot demonstration of the rich physical information that can be obtained by combining ground-based integral-field UV spectroscopy with deep JWST/NIRCam emission-line imaging. This approach can be applied to a growing number of LRDs with similarly rich ancillary datasets.

Throughout this paper, we adopt the AB magnitude system and a $\Lambda$CDM cosmology with \citet{Planck2020} parameters, i.e., $\Omega_m = 0.315$ and $h = H_0/(100\,{\rm km\,s^{-1}\,Mpc^{-1}}) = 0.673$.

\section{Observations}
\label{sec:obs}

In this study, we use VLT/MUSE integral-field data from the final official data release of the MUSE Hubble Deep Field South surveys \citep{Bacon2017,Bacon2023}. The source lies in the MOSAIC portion of the survey \citep[see][]{Bacon2023}, which reaches a depth of 10 hr exposure time. The science-ready products, including the full data cube, extracted 1D spectrum, and emission-line maps, were obtained through the \texttt{AMUSED} web interface\footnote{\href{https://amused.univ-lyon1.fr/project/UDF/}{https://amused.univ-lyon1.fr/project/UDF/}}. As detailed in Appendix~\ref{app:astrometry}, an astrometric correction was applied to align the MUSE World Coordinate System (WCS) with the NIRCam imaging mosaics; after this correction, the relative astrometric precision reaches a median of zero and a standard deviation of $\approx 0.07''$.

We also use NIRCam imaging data obtained from the JADES survey \citep{Rieke2023,Eisenstein2026}, together with NIRCam medium-band imaging from the JEMS survey \citep{Williams2023}. All NIRCam mosaics used in this study are taken from the latest JADES data release, DR5 \citep{Johnson2026,Robertson2026}. The images have a drizzle scale of $0.03''$ pix$^{-1}$ and are registered to the Gaia DR3 WCS reference frame \citep{Gaia2021}. We adopt JADES PSF models constructed for the region containing the source. These models are based on \textsc{webbpsf} \citep{Perrin2014}, while carefully accounting for data-reduction effects, and have been validated against point sources observed in JADES \citep{Ji2024}. They are consistent with the observed point sources at the $\lesssim 3\%$ level in the inner core, while retaining the high S/N that would otherwise be difficult to achieve with conventional empirical PSF models because of the limited number of bright, isolated stars in GOODS-S.

\section{UV Emission-line Properties from MUSE}
\label{sec:muse}

\subsection{The double-peaked Ly$\alpha$ profile and tentative N\,{\sc v} detection}

\begin{figure*}
\centering
    \includegraphics[width=0.87\textwidth]{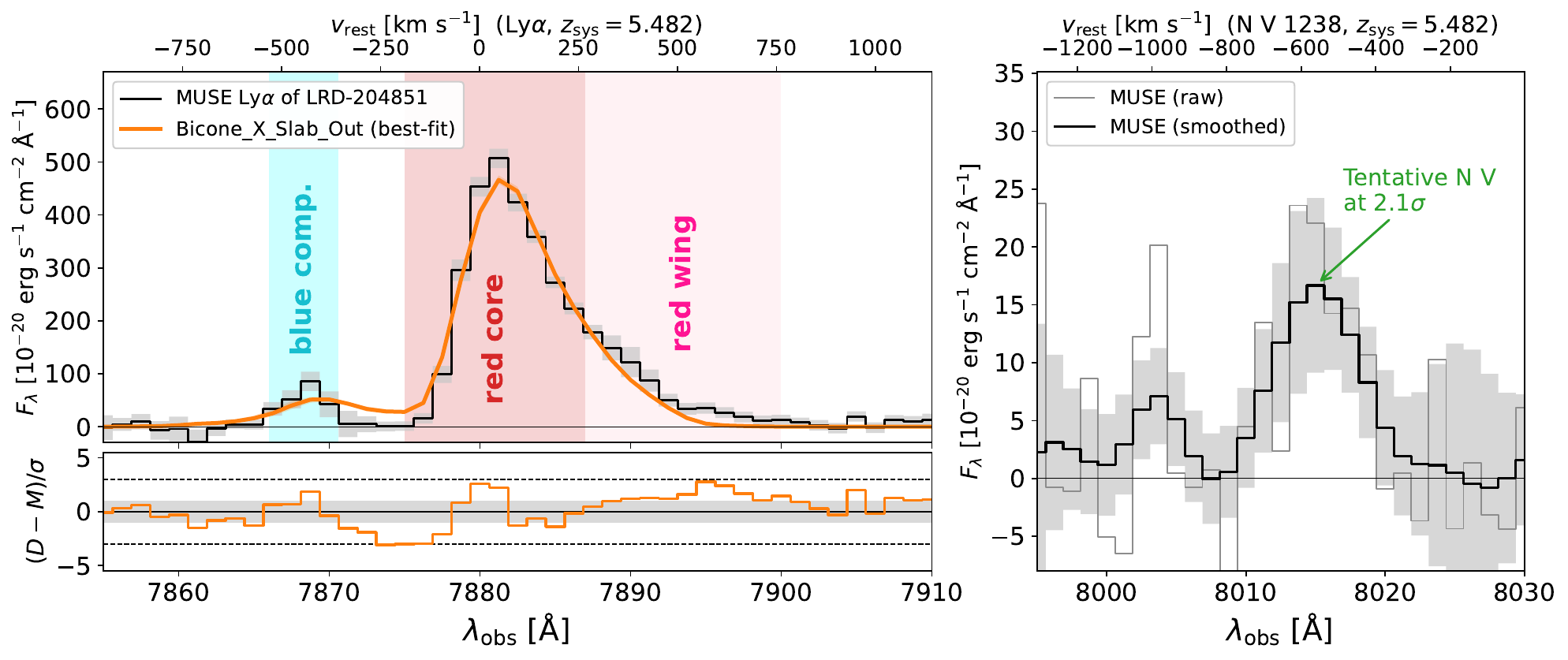}
    \caption{\textbf{Left:} MUSE spectrum (black) of the Ly$\alpha$ region with the \texttt{Bicone\_X\_Slab\_Out} best-fit model (orange) overlaid. The cyan, red, and pink bands mark the disjoint windows used to build the blue-component, red-core, and red-wing narrow-band maps (see Section \ref{sec:muse_resolved}). The lower sub-panel shows the fit residuals with $\pm 1\sigma$ (grey band) and $\pm 3\sigma$ (dashed) reference levels. The full posterior is shown in Appendix~\ref{app:geometries}. \textbf{Right:} MUSE spectrum around N\,{\sc v}\,$\lambda$1238, with the raw flux (light grey) overlaid by a 1.5-pixel (i.e., $\approx$1.8\AA) Gaussian-smoothed version (black) and its propagated $\pm 1\sigma$ band. The arrow marks the tentative $2.1\sigma$ N\,{\sc v}\,$\lambda$1238 detection. }
    \label{fig:muse_line}
\end{figure*}

As Figure \ref{fig:muse_line} shows, the MUSE observations reveal a multi-component Ly$\alpha$
emission line and a tentative detection of N\,{\sc v}
$\lambda 1238$ in LRD-204851.  In the official
catalog from the MUSE HUDF team \citep{Bacon2023}, the
Ly$\alpha$ line is decomposed into two Gaussian components:
a narrow blue peak at
$\lambda_{\rm obs}=7868.6$\,\AA\ and a broader red
peak at $\lambda_{\rm obs}=7880.9$\,\AA.  The combined Ly$\alpha$ blend has total flux
$F_{\rm Ly\alpha}=(4.1\pm 0.1)\times 10^{-17}\,\rm erg\,s^{-1}\,cm^{-2}$
(rest-frame EW $\simeq\!144$\,\AA, $L_{\rm Ly\alpha}\simeq
1.5\times 10^{43}\,\rm erg\,s^{-1}$), of which $\sim\!6\%$
sits in the blue component
($F_{\rm Ly\alpha,blue}=(2.4\pm 0.5)\times 10^{-18}\,\rm erg\,s^{-1}\,cm^{-2}$).
The tentative N\,{\sc v}\,$\lambda 1238$ line is reported at
$\lambda_{\rm obs}=8015.3$\,\AA{} with S/N\,$=2.1$.  
We note that N\,{\sc v} is a resonance doublet, with the redder $\lambda1242$ component intrinsically a factor of $\sim2$ weaker than the $\lambda1238$ one in the optically thin limit, set by the 2:1 ratio of oscillator strengths \citep{Morton2003}. Given the modest S/N $\simeq2.1$ of the $\lambda1238$ line, the $\lambda1242$ component is therefore expected to be undetected. As a consistency check, we also independently re-extract the 1D spectrum from the WCS-corrected cube in the AMUSED Ly$\alpha$ and N\,{\sc v} emission-line segmentation apertures (with local-sky subtraction) and integrate the line fluxes over the AMUSED bandpasses. Our measurements and the \citet{Bacon2023} catalog Ly$\alpha$ and N\,{\sc v} fluxes, as well as the blue/red peak velocities, agree with each other at the $\lesssim10\%$ level. We adopt the published \citet{Bacon2023} values throughout.

Relative to the systemic redshift $z_{\rm sys}=5.482$ (based on the \mbox{[\ion{O}{3}]} emission line from a NIRSpec high-resolution spectrum presented in \citealt{Matthee2026}), the red Ly$\alpha$ peak sits at $\Delta v=-38\,\rm km\,s^{-1}$, while the blue Ly$\alpha$ peak and the tentative N\,{\sc v}\,1238 line are blueshifted by
$\Delta v_{\rm Ly\alpha,blue}=-431\pm 9\,\rm km\,s^{-1}$ and $\Delta v_{\rm NV}=-467\pm 241\,\rm km\,s^{-1}$, respectively (Figure \ref{fig:muse_line}). The blue Ly$\alpha$ peak and N\,{\sc v} are therefore consistent with being at the same velocity within uncertainties, providing additional support for the otherwise tentative N\,{\sc v} detection.

The detection of a blueshifted Ly$\alpha$ peak at $z\approx5.5$ is itself notable: at this redshift, the IGM neutral \ion{H}{1} fraction typically extinguishes such peaks along an average sightline
\citep{Madau1995, Inoue2014, Laursen2011, Bosman2022}. Confirmed blue-peak detections at such high redshift have been attributed to large local ``proximity zones'' around the source, where the source's own ionizing flux, e.g. from a quasar, exceeds the UV background and lowers the local IGM neutral fraction. The most notably example is COLA1 at $z=6.59$, whose blue peak is consistent with a $\sim0.7$\,Mpc ionized bubble carved by its own high Lyman continuum (LyC) escape fraction \citep[e.g.,][]{Hu2016,Matthee2018, Torralba2024}.  Local IGM underdensities have also been proposed as a theoretical alternative \citep[e.g.,][]{Park2026}.  

We assess the proximity-zone possibility for the blue Ly$\alpha$ peak observed in LRD-204851.  Using the measured UV continuum magnitude ($M_{1500}=-18.82$) and
UV slope ($\beta=-1.38$) from \citet{lya2026}, we infer a LyC luminosity of $L_{912}\simeq1.1\times10^{28}$\,erg\,s$^{-1}$\,Hz$^{-1}$. We then estimate $R_{\rm eq}$, which denotes the distance from the source where the photoionization rates of the source $\Gamma_{\rm source}(r)$ and of the cosmic UV background $\Gamma_{\rm bg}$  would be equal for purely geometric dilution in the absence of attenuation. 
Following the $R_{\rm eq}$ formalism of \citet{Zhu2023}, and
adopting an ionizing-spectrum slope $\alpha_\nu^{\rm ion}=1.5$ and a
$\Gamma_{\rm bg}=5\times10^{-13}$\,s$^{-1}$ representative of
$z\sim5.5$ \citep[e.g.,][]{Becker2013, Bosman2022}, this corresponds
to a proximity-zone radius $R_{\rm eq}\simeq0.2$\,Mpc.  This
is substantially smaller than the $\simeq0.7$\,Mpc Hubble-flow distance corresponding to a $\Delta v_{\rm Ly\alpha,blue}\simeq-430$\,km\,s$^{-1}$ photon, so a proximity zone alone cannot account for the observed blue Ly$\alpha$ peak.

\subsection{Ly$\alpha$ profile modeling}
\label{sec:lya_rt}
Our calculation above indicates that the proximity zone can only provide a supporting transmission channel that allows blue Ly$\alpha$ photons in LRD-204851 to remain visible through the surrounding CGM and nearby IGM. The observed blue Ly$\alpha$ emission should therefore arise primarily from local radiative-transfer processes. This interpretation is further motivated by the Ly$\alpha$ profile itself: a broad red peak near systemic accompanied by a blueshifted component is characteristic of Ly$\alpha$ escape through a partially ionized, kinematically structured neutral medium \citep[e.g.,][]{Verhamme2006,Hansen2006,Dijkstra2006}.

\begin{figure*}
    \includegraphics[width=1\textwidth]{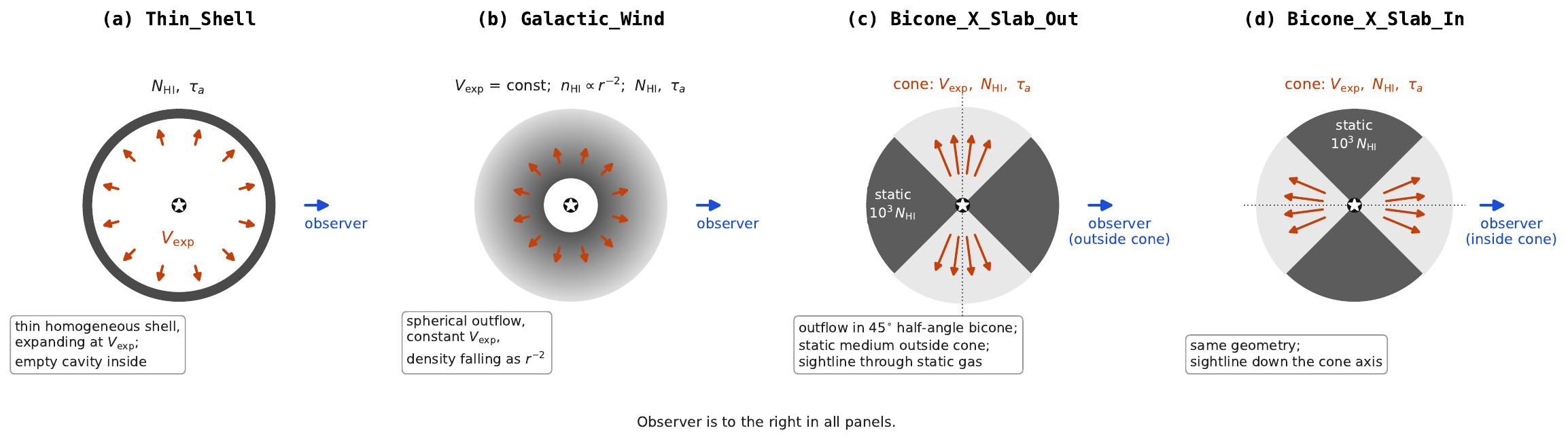}
    \caption{Schematic of the four \texttt{zELDA} gas configurations used to model the Ly$\alpha$ profile (Section~\ref{sec:lya_rt}), with the observer located to the right in each panel. Grey regions indicate neutral gas, with darker shading denoting higher column density, and arrows indicate the radial outflow velocity $V_{\rm exp}$. Panels show (a) \texttt{Thin\_Shell}, (b) \texttt{Galactic\_Wind}, (c) \texttt{Bicone\_X\_Slab\_Out}, and (d) \texttt{Bicone\_X\_Slab\_In}. The two biconical models have the same gas geometry and differ only in viewing angle.}
    \label{fig:illustration}
\end{figure*}

We therefore model the Ly$\alpha$ profile with \texttt{zELDA} \citep{GurungLopez2019,GurungLopez2022}, which provides precomputed Ly$\alpha$ profiles from the Monte Carlo radiative-transfer code \texttt{LyaRT} \citep{Orsi2012}. We consider the four gas configurations implemented in \texttt{zELDA}, illustrated in Figure~\ref{fig:illustration}. In all of them, a central point source emits Ly$\alpha$ into an isothermal ($T=10^{4}$~K) neutral medium. The fit has four free parameters, defined in the same way for all configurations: 
\begin{itemize}
    \item $z_{{\rm Ly}\alpha}$, the redshift of the central source, i.e.\ the redshift at which the intrinsic Ly$\alpha$ line is emitted before radiative transfer; 
    \item $V_{\rm exp}$, the radial velocity of the outflowing neutral gas; 
    \item $N_{\rm HI}$, the H\,{\sc i} column density of the outflowing gas along a radial path from the source; 
    \item $\tau_a$, the dust absorption optical depth of the same gas at the Ly$\alpha$ wavelength. 
\end{itemize}
The configurations differ in the spatial distribution and kinematics of the gas (Figure \ref{fig:illustration}). These are: (1) \texttt{Thin\_Shell}, a homogeneous spherical shell surrounding an empty cavity and expanding at $V_{\rm exp}$ \citep{Verhamme2006}; (2) \texttt{Galactic\_Wind}, a spherical outflow at constant $V_{\rm exp}$ with density decreasing as $n_{\rm HI}\propto r^{-2}$ outside an empty central cavity \citep{Orsi2012}; (3) \texttt{Bicone\_X\_Slab\_Out}, in which the outflow is confined to a bicone of half-opening angle $45^{\circ}$ and surrounded by a static medium with a column density $10^{3}$ times higher, with the observer located outside the cone; and (4) \texttt{Bicone\_X\_Slab\_In}, the same geometry viewed from inside the cone, i.e.\ down the cone axis.

Among the four gas geometries available in \texttt{zELDA}, the \texttt{Bicone\_X\_Slab\_Out} geometry is strongly preferred. It is the only model that simultaneously reproduces the near-systemic red peak and the blueshifted blue companion of LRD-204851 (Figure~\ref{fig:muse_line}), and it is favored over each of the other three geometries by the Bayesian evidence at the ``decisive'' level ($\Delta\ln Z \geq 8$ for all alternatives; see Appendix~\ref{app:geometries} for the detailed four-geometry comparison). The marginalized parameters from its full posterior are
$z_{\rm Ly\alpha}=5.4792$
($\Delta v \simeq -143\,{\rm km\,s^{-1}}$ relative to $z_{\rm sys}$ derived from \mbox{[\ion{O}{3}]}),
$V_{\rm exp}\simeq 110\,{\rm km\,s^{-1}}$,
$\log (N_{\rm HI}/{\rm cm}^{-2})=21.27\pm0.03$, and
$\tau_{\rm a}\simeq0.30$. In this picture, the dominant red peak emerges from photons that back-scatter through the optically thick neutral slab, while the blue feature is produced by the much weaker cone-leakage path. This configuration naturally accounts for both the $\sim\!400\,{\rm km\,s^{-1}}$ separation between the two Ly$\alpha$ peaks and the profile of the red-component line wing. The inferred high $N_{\rm HI}$ and modest $V_{\rm exp}$ together imply an obscured, gas-rich circumnuclear environment with a biconical cavity oriented away from the line of sight.

While the integrated profile modeling clearly favors \texttt{Bicone\_X\_Slab\_Out} over the other three \texttt{zELDA} families, the gas geometries explored here are, by construction, idealized. Each assumes a smooth, axisymmetric, homogeneous neutral medium with a single bulk outflow velocity and a Gaussian intrinsic Ly$\alpha$ source. The real ISM in galaxies is almost certainly more complex: multiphase, clumpy, and turbulent, with a distribution of cloud column densities and velocities along any sightline \citep{Gronke2017}. Shell-type fits are also known to carry strong internal degeneracies, particularly among $V_{\rm exp}$, gas temperature, and the intrinsic Ly$\alpha$ line width, which are only partially broken when external constraints are available \citep{LiGronke2022}. Moreover, very clumpy geometries asymptotically converge to the same family of profiles as a smooth expanding shell once the number of clouds along the sightline becomes large \citep{Gronke2016}. We therefore make no claim that the neutral gas in LRD-204851 takes the exact form of a thin slab pierced by a biconical cavity. Rather, our results only suggest that, among the four characteristic geometries available in \texttt{zELDA}, the integrated profile is most consistent with this configuration, and that the marginalized parameters provide a useful physical scale for the obscuring column and outflow speed. The spatial distribution of the line-emitting gas, examined in Section~\ref{sec:muse_resolved}, provides a complementary and largely independent test: under the favored geometry, the blue cone-leakage component should be spatially displaced from the resonantly back-scattered red core along the cavity axis, while the tentative N\,{\sc v} feature would align with the same direction as the blue Ly$\alpha$ component.

We further note that the slow expansion velocity inferred here is independently corroborated by the NIRSpec R2700 spectrum of LRD-204851 reported by \citet{Matthee2026}: their line-profile fit detects a blueshifted \mbox{H$\alpha$} absorber at $\Delta v \approx -80$\,km\,s$^{-1}$, kinematically consistent with the $V_{\rm exp} \approx 110$\,km\,s$^{-1}$ of the Ly$\alpha$-emitting envelope inferred above.  Since the Ly$\alpha$ and Balmer absorbers probe different transitions (n=1 and n=2, respectively) of the same partially ionized neutral medium, the agreement provides an independent check that the central source of LRD-204851 is enshrouded by a slowly expanding gas envelope or clumpy medium of the kind required by our \texttt{zELDA} modeling.

\subsection{A spatially resolved view of the Ly$\alpha$ components and N\,{\sc v}}
\label{sec:muse_resolved}

We use the MUSE data cube to examine the spatial distribution of the Ly$\alpha$ and tentative N\,{\sc v} emission in LRD-204851. The MUSE HUDF release provides narrow-band emission-line maps for both the
integrated Ly$\alpha$ blend and N\,{\sc v}\,$\lambda1238$ \citep{Bacon2023}, and we adopt the latter directly. To spatially separate the multi-component Ly$\alpha$ profile, however, we construct
three disjoint Ly$\alpha$ narrow-band maps over the wavelength windows illustrated in Figure~\ref{fig:muse_line}: the blue component
($7866.0$--$7870.6$\,\AA), the red core ($7875.0$--$7887.0$\,\AA), and the red wing ($7887.0$--$7900.0$\,\AA). The maps are built using
the MUSE Python Data Analysis Framework (\texttt{MPDAF}; \citealt{Piqueras2019}), following the same recipe used by the MUSE HUDF team \citep{Bacon2023}. We have verified that applying this recipe over the full Ly$\alpha$ window reproduces their published narrow-band map to better than one percent within the segmentation aperture.

\begin{figure*}
    \includegraphics[width=1\textwidth]{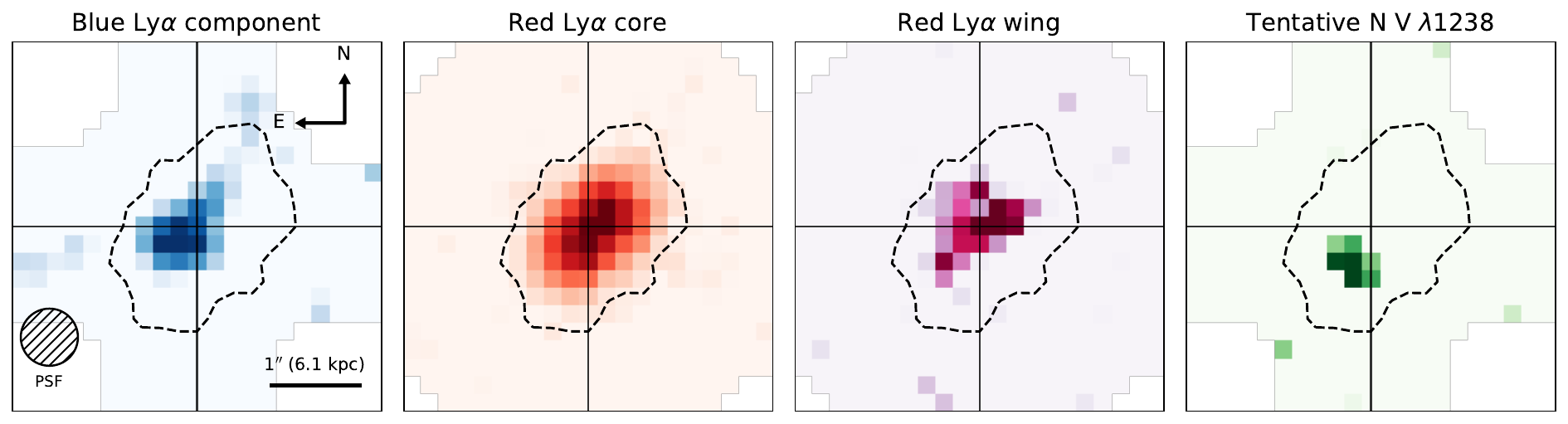}
    \caption{MUSE narrow-band maps of the multi-component Ly$\alpha$ emission and the tentative N\,{\sc v}\,$\lambda1238$ detection in LRD-204851. From left to right: the blue Ly$\alpha$ component, the red Ly$\alpha$ core, the red Ly$\alpha$ wing, and the N\,{\sc v}\,$\lambda1238$ narrow-band map. Per-panel color stretches saturate at the 99.5th percentile and use a lower limit of $1\sigma$ for the displayed map, so structures below the noise level fade to white. The black dashed contour shows the boundary of the Ly$\alpha$ segmentation map from the MUSE HUDF release and is identical in every panel; the thin black crosshairs mark the NIRCam/F460M continuum centroid.}
    \label{fig:muse_nb}
\end{figure*}

Figure~\ref{fig:muse_nb} shows the four MUSE narrow-band maps side by side. The red Ly$\alpha$ core map is centered on the rest-frame optical continuum (F460M) center to within one MUSE pixel ($0.2''$), and the red Ly$\alpha$ wing map appears to trace the same overall source. By contrast, the blue Ly$\alpha$ component appears displaced to the southeast of the F460M centroid: a peak-pixel centroid measurement gives
$\Delta\alpha\cos\delta=-0.34''$ and $\Delta\delta=+0.07''$, corresponding to a projected separation of $|\Delta r|=0.34''$ ($\simeq2.1$\,kpc at $z=5.482$). To assess the significance of this offset, we injected $10^{4}$ synthetic point sources with matched blue-Ly$\alpha$ flux at random empty-sky positions within the $2\times2$ arcmin$^2$ MUSE mosaic cutout around LRD-204851. We then recovered each centroid using the same centroiding algorithm. The recovered offsets have a mean consistent with zero and a per-axis standard deviation of $\sigma\simeq0.10''$. Among the $10^{4}$ realizations, 155 produced a recovered $|\Delta r|\geq0.34''$, corresponding to a one-sided $p$-value of $p\simeq0.016$ for the null hypothesis that the blue Ly$\alpha$ centroid coincides with the red Ly$\alpha$ core. The observed offset is therefore unlikely to be a spurious centroiding artifact. 

The narrow-band map constructed for the tentative N\,{\sc v}\,$\lambda1238$ detection also shows a peak along the same south-eastern direction from the F460M continuum centroid as the blue Ly$\alpha$ peak, as shown in the rightmost panel of  Figure~\ref{fig:muse_nb}. The fact that the N\,{\sc v} peak both lies along the same direction as the blue Ly$\alpha$ component and shares a consistent blueshift relative to $z_{\rm sys}$ in the integrated spectrum (Section~\ref{sec:lya_rt}) provides circumstantial support for the otherwise tentative N\,{\sc v}\,$\lambda1238$ detection, although higher-S/N follow-up will ultimately be required to confirm it. We emphasize that the tentative N\,{\sc v} feature is not required for the main morphological conclusion of this work. If future higher-S/N observations do not confirm the N\,{\sc v} feature, the alignment between the blue Ly$\alpha$ component and the south-eastern \mbox{[\ion{O}{3}]}/UV structure would remain unchanged, but the evidence for a very hard ionizing radiation field at this location would be weakened. We therefore treat N\,{\sc v} as suggestive supporting evidence rather than as a necessary pillar of the interpretation.

The spatial picture from MUSE alone, therefore, is consistent with one in which the red Ly$\alpha$ core traces material spatially co-located with the LRD's optical continuum, while the blue Ly$\alpha$ component and the N\,{\sc v}\,$\lambda 1238$ emission both peak $\sim\!2$\,kpc to the south-east. This is qualitatively the spatial signature expected if the blue Ly$\alpha$ component represents Ly$\alpha$ photons leaking through a low-column-density channel offset from the source, while
the red core is the resonantly back-scattered emission from the optically thick neutral medium surrounding it, the same picture favored by the integrated profile analysis in Section~\ref{sec:lya_rt}. We caution, however, that the MUSE data alone cannot exclude the offset blue-Ly$\alpha$ and N\,{\sc v} emission arising in a physically distinct structure (e.g., a close companion) rather than from cone leakage connected to the central engine; we return to this possibility in Section~\ref{sec:discussion}.

\section{Optical Emission-line Maps from NIRCam Imaging}
\label{sec:nircam}

Rest-frame optical emission lines provide powerful diagnostics of ionized gas, and at $z\approx5.5$ these lines fall within the NIRCam wavelength coverage. In this section, we use NIRCam imaging to construct continuum-subtracted maps, which provide a sub-kpc-resolution view of \mbox{[\ion{O}{3}]}, H$\beta$, and H$\alpha$+\mbox{[\ion{N}{2}]}. LRD-204851 is particularly well suited for this analysis because it lies near the HUDF in GOODS-S, where JADES \citep{Eisenstein2026} and JEMS \citep{Williams2023} together provide deep NIRCam imaging in 14 broad- and medium-band filters at the source position, spanning F090W to F480M. This dense coverage makes it possible to isolate individual rest-frame optical emission lines through on-band photometry while constraining the underlying continuum from line-free anchors.

\subsection{Construction of the optical line maps}
\label{sec:linemap_construction}

At $z=5.482$, H$\alpha$+\mbox{[\ion{N}{2}]} falls at $\lambda_{\rm obs}\approx4.25$--$4.27\,\mu$m, where F430M provides the cleanest medium-band coverage, with a transmission of $T\sim0.50$ at the line wavelengths\footnote{These lines are also covered by F410M ($T\sim0.4$, lower than that of F430M) and by the broad-band F444W, but
with lower line-to-continuum contrast. We therefore use F430M as the on-band tracer of H$\alpha$+\mbox{[\ion{N}{2}]}.}. We do not attempt to separate H$\alpha$ from \mbox{[\ion{N}{2}]},
nor to estimate the \mbox{[\ion{N}{2}]} contribution explicitly in
this work.  While the existing NIRSpec/R1000
spectrum \citep{Juodzbalis2026} shows very weak \mbox{[\ion{N}{2}]} emission from the LRD's optical continuum peak (slit position shown in Figure~\ref{fig:target_info}), we caution that LRD-204851 exhibits complex spatial structure, so it remains unclear whether \mbox{[\ion{N}{2}]} is similarly weak across the entire galaxy. Resolving this would require NIR IFU observations.

For \mbox{[\ion{O}{3}]} and H$\beta$, separation using NIRCam imaging alone is generally not feasible, since the two lines are typically covered by the same filter with similar throughputs. At the redshift of LRD-204851, however, both lines fall within F335M and F356W but with markedly different filter throughputs: at $\lambda_{\rm obs}({\rm H}\beta)=3.15\,\mu$m, the F335M transmission is $T=0.07$, compared to $T=0.34$ in F356W, a factor of $\sim5$ difference, while the \mbox{[\ion{O}{3}]} doublet falls on the central plateau of both filters with similar transmissions. This asymmetry makes F335M a nearly pure \mbox{[\ion{O}{3}]} tracer and F356W a mixed \mbox{[\ion{O}{3}]}+H$\beta$ tracer, allowing us to separate the \mbox{[\ion{O}{3}]} and H$\beta$ maps.

Because the rest-frame UV and optical continua of LRD-204851 are described by distinct power laws that meet around the Balmer break (Figure~\ref{fig:target_info}), we estimate the continuum using only the remaining filters that sample $\lambda_{\rm rest}\gtrsim4000$~\AA, where a single power law provides an adequate local description of the SED. We therefore adopt F277W, F460M, and F480M as line-free continuum anchors. We PSF-match all bands used in the analysis to the F480M resolution using kernels generated with \texttt{pypher} \citep{Boucaud2016}. For each pixel, we fit a power-law continuum, subtract the continuum prediction from each on-band filter, and convert the residual surface brightness to integrated line flux per pixel using the NIRCam transmission curves \citep{Pascual2007,Ion1}.

\subsection{Spatial distribution of the optical line emission}
\label{sec:nircam_linemap}

\begin{figure*}
\centering
    \includegraphics[width=0.97\textwidth]{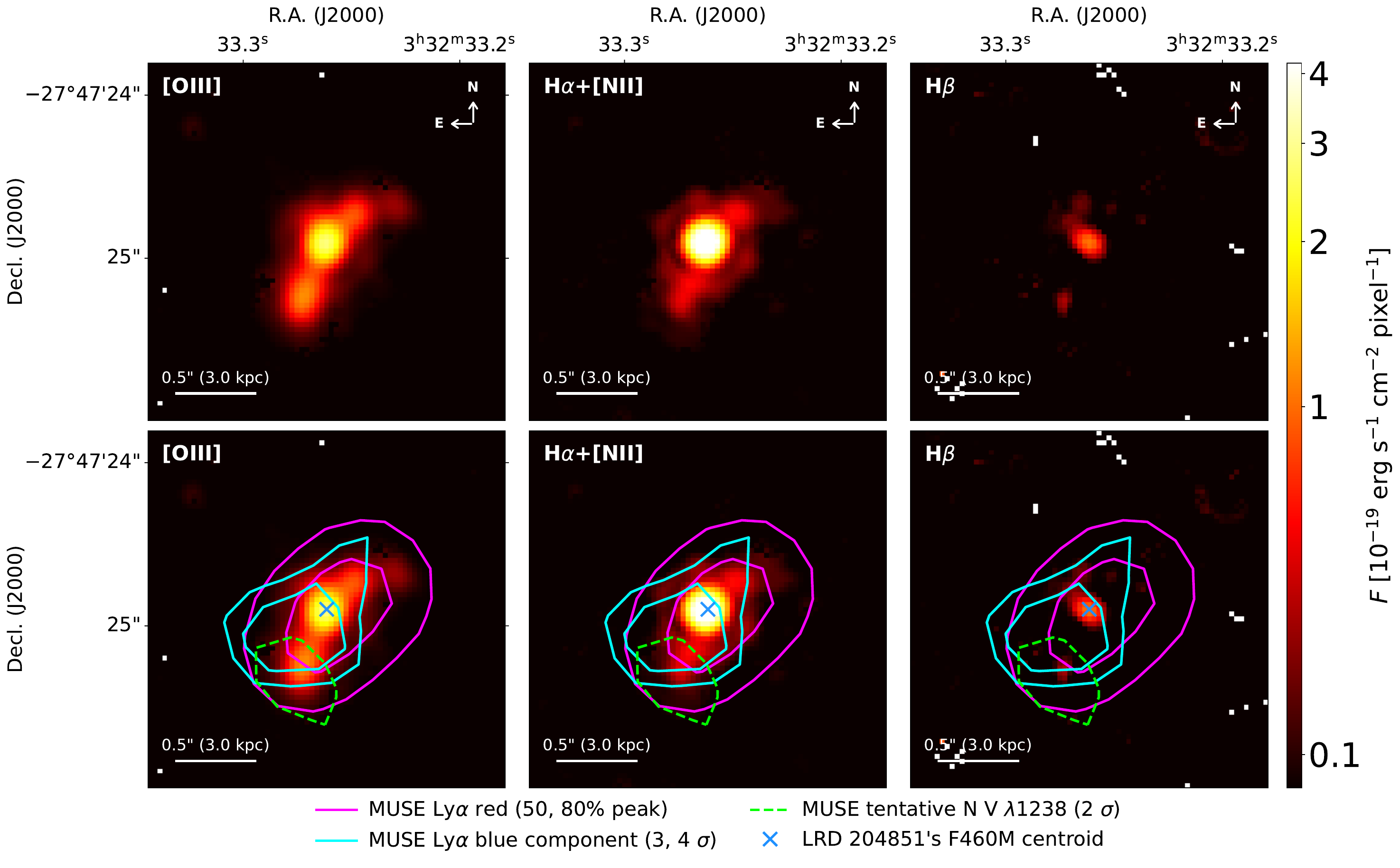}
    \caption{Optical emission-line maps of LRD-204851 constructed using NIRCam imaging.  From left to right: \mbox{[\ion{O}{3}]}, H$\alpha$+\mbox{[\ion{N}{2}]}, and H$\beta$. The top row shows the maps alone; the bottom row overlays the MUSE red (magenta) and blue (cyan) Ly$\alpha$ component contours and the tentative \mbox{N\,{\sc v}}\,$\lambda1238$ contour (green dashed), with the NIRCam/F460M continuum centroid marked by the blue $\times$.  All panels share a common color scale of line flux per pixel.}
\label{fig:lya_NV_on_3lines}
    \label{fig:nircam_linemaps}
\end{figure*}

\begin{figure}
    \centering
    \includegraphics[width=0.87\linewidth]{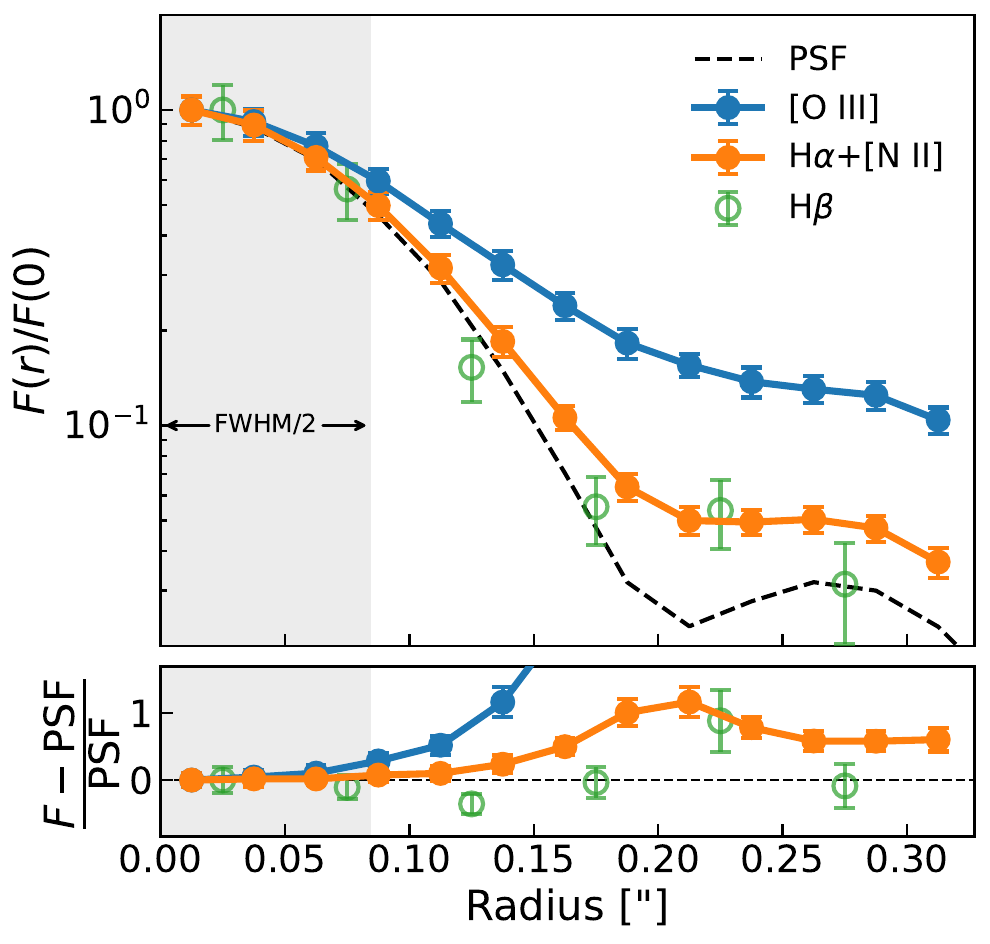}
    \caption{Azimuthally averaged radial surface-brightness profiles of the optical emission-line maps in LRD-204851 (Figure~\ref{fig:lya_NV_on_3lines}), normalized by the central-annulus value. The black dashed curve shows the radial profile of the PSF, and the gray shaded region marks the scale of the PSF's ${\rm FWHM}/2$. The bottom panel shows the relative difference from the PSF. The H$\alpha$+\mbox{[\ion{N}{2}]} profile tracks the PSF within $r \lesssim 0.15''$, while the \mbox{[\ion{O}{3}]} profile lies above the PSF, indicating a more extended light distribution than H$\alpha$+\mbox{[\ion{N}{2}]}.}
    \label{fig:radial}
\end{figure}

Figure~\ref{fig:lya_NV_on_3lines} shows the resulting
\mbox{[\ion{O}{3}]}, H$\alpha$+\mbox{[\ion{N}{2}]}, and H$\beta$ maps.
These rest-frame optical lines reveal distinct morphological features.
The H$\alpha$+\mbox{[\ion{N}{2}]} map is dominated by a bright central
source whose inner surface-brightness profile is consistent with the
PSF within $\sim0.15''$ (Figure~\ref{fig:radial}). Around this
central source, the H$\alpha$+\mbox{[\ion{N}{2}]} emission also shows a thin, fainter
elongated structure extending along the southeast-to-northwest axis. 

The \mbox{[\ion{O}{3}]} map similarly
shows a central clump at the F460M centroid, but this clump is resolved
and inconsistent with being PSF-like (Figure~\ref{fig:radial}). \mbox{[\ion{O}{3}]} also displays an elongated structure along the same
southeast-to-northwest axis traced by H$\alpha$+\mbox{[\ion{N}{2}]} and
the blue UV continuum, with a bright clump-like feature at the
southeastern end of this structure that is visible in the UV
continuum (Figure~\ref{fig:target_info}). The H$\beta$ map has
relatively low S/N, but emission is detected both at the center
of the LRD and in the southeastern \mbox{[\ion{O}{3}]} clump region.

The bottom row of Figure~\ref{fig:lya_NV_on_3lines} overlays the
MUSE Ly$\alpha$ components and the tentative \mbox{N\,{\sc v}} contours on the optical line maps. The blue Ly$\alpha$ contour is offset to the south-east of the \mbox{F460M} continuum centroid, along the same direction as the elongated structure traced by the UV continuum, H$\alpha$+\mbox{[\ion{N}{2}]}, and \mbox{[\ion{O}{3}]}. Because the MUSE PSF ($\mathrm{FWHM}\approx0.7''$) is much coarser than the NIRCam resolution, this agreement is robust only as an overall south-east alignment. The data do not constrain how closely the blue Ly$\alpha$ emission follows the optical structure in detail: they do not constrain either how thin and collimated it is or the degree to which the two are spatially coincident beyond the MUSE resolution.  The  tentative \mbox{N\,{\sc v}} contour appears confined to the bright southeastern \mbox{[\ion{O}{3}]} clump region.

\begin{figure}
    \centering
    \includegraphics[width=1\linewidth]{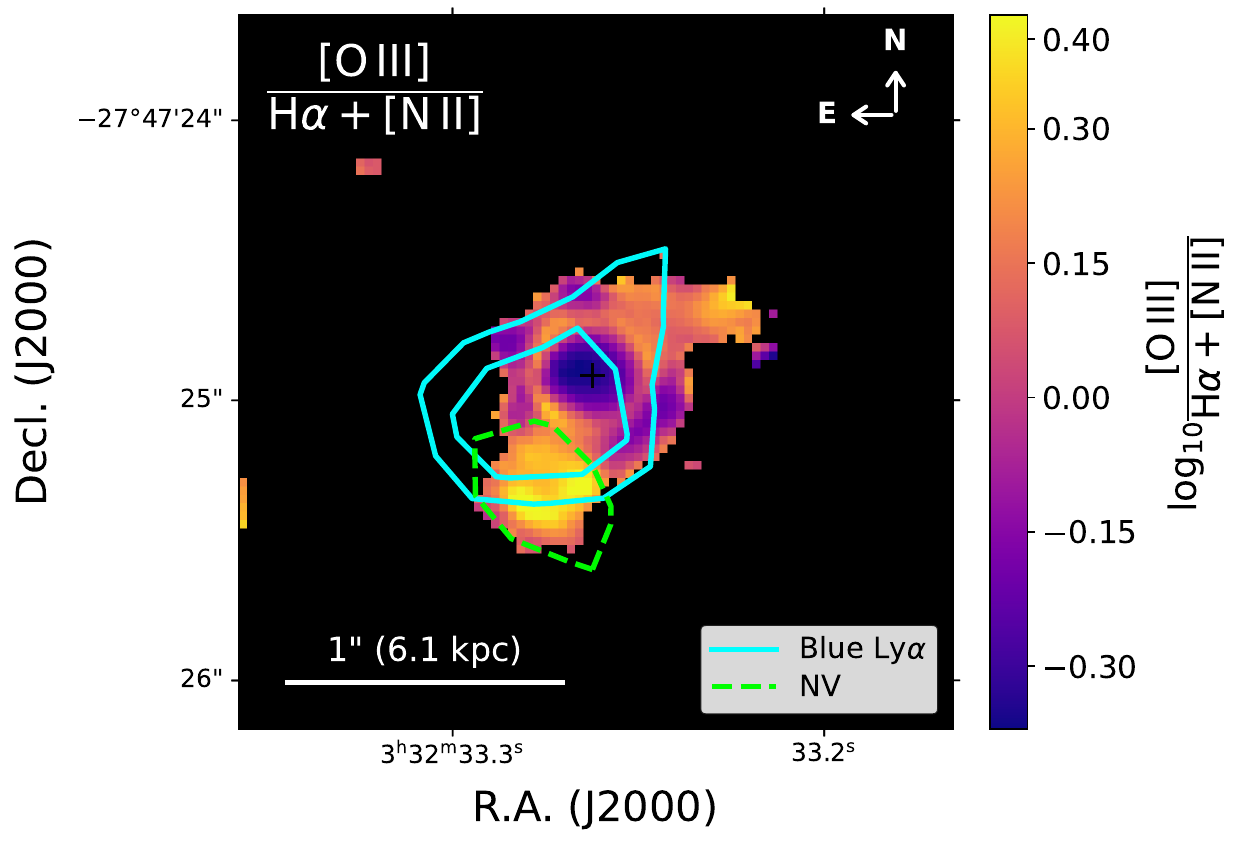}
    \caption{Map of the line ratio $\log_{10}(\mbox{[\ion{O}{3}]}\,/\,{\rm H}\alpha\!+\!\mbox{[\ion{N}{2}]})$ in LRD-204851, derived from the line maps shown in Figure~\ref{fig:lya_NV_on_3lines}. Pixels are masked where either line has S/N $<3$. Overlaid contours show the MUSE blue Ly$\alpha$ component and the tentative N\,{\sc v}\,$\lambda1238$ emission, as in Figure~\ref{fig:lya_NV_on_3lines}.}
    \label{fig:line_ratio}
\end{figure}

The corresponding
$\mbox{[\ion{O}{3}]} / (\text{H}\alpha + \mbox{[\ion{N}{2}]})$
line-ratio map (Figure~\ref{fig:line_ratio}), a proxy for the
ionization state of the photoionized gas, with higher values
indicating more highly ionized gas \citep[e.g.,][]{Baldwin1981,Kewley2019}, further shows that this southeastern region exhibits the highest $\mbox{[\ion{O}{3}]} / (\text{H}\alpha + \mbox{[\ion{N}{2}]})$ ratios in the source, while the central LRD continuum peak has a much lower (0.7~dex) ratio. We caution that the broad-band-derived [\mbox{O\,{\sc iii}}]/(\mbox{H$\alpha$}+[\mbox{N\,{\sc ii}}]) ratio at the LRD continuum center is significantly diluted by the broad \mbox{H$\alpha$} emission line.  At the LRD continuum peak, the broad component carries $\approx 70\%$ of the total \mbox{H$\alpha$} flux, with the narrow component contributing only $F(\mathrm{H\alpha,\,narrow}) \approx 5\times10^{-18}\, \mathrm{erg\,s^{-1}\,cm^{-2}}$ \citep{Juodzbalis2026}.  Removing the broad-line contribution and using only the narrow \mbox{H\,$\alpha$} flux yields $\log_{10}F([\mbox{O\,{\sc iii}}]\,5007)/F(\mathrm{H\alpha,\,narrow}) \approx 0.34$ at the LRD centre, comparable to the highest values seen in the south-eastern clump on the broad-band ratio map. The narrow-line ionization state of LRD-204851 is therefore
high across the entire source, with the apparent radial gradient in
Figure~\ref{fig:line_ratio} driven primarily by the spatial concentration of the broad \mbox{H$\alpha$} BLR rather than by a genuine change in the ionization state of the narrow-line gas. Notably, the tentative \mbox{N\,{\sc v}} emission, which requires a very hard ionizing radiation field (the ionization potential to produce $\mathrm{N}^{4+}$ is ${\sim}77$~eV, compared to ${\sim}35$~eV for $\mathrm{O}^{2+}$; see \citealp{treiber2025}), peaks on top of this same southeastern high-ionization region, lending additional
confidence to the otherwise tentative \mbox{N\,{\sc v}} detection from MUSE.

\subsection{Continuum-emission analysis of the southeastern clump} \label{sec:cont_sed}
To further investigate the nature of the southeastern clump, we examine its continuum SED using the NIRCam imaging. We focus on this region because it shows the strongest evidence for an additional energetic, non-stellar contribution, including its elevated $\mbox{[\ion{O}{3}]} / (\text{H}\alpha + \mbox{[\ion{N}{2}]})$ ratio and its spatial association with the blueshifted Ly$\alpha$ and tentative \mbox{N\,{\sc v}} emission. 

We measure photometry within a circular aperture of radius $0.15''$ centered on the clump, approximately $0.4''$ south-east of the optical-continuum centroid. We use the six filters free of strong emission-line contamination at $z=5.482$: F150W, F182M, F200W, F210M, F460M, and F480M. We fit these measurements with \texttt{Prospector} \citep{Johnson2021} with redshift fixed. We adopt the same SED assumptions as used in \citet{Ji2024}, namely a nine-bin continuity star-formation history, a two-component dust model \citep{Charlot2000}, and self-consistent \texttt{CLOUDY} nebular emission \citep{Byler2017} with free gas-phase metallicity and ionization parameter. 

\begin{figure}
    \centering
    \includegraphics[width=1\linewidth]{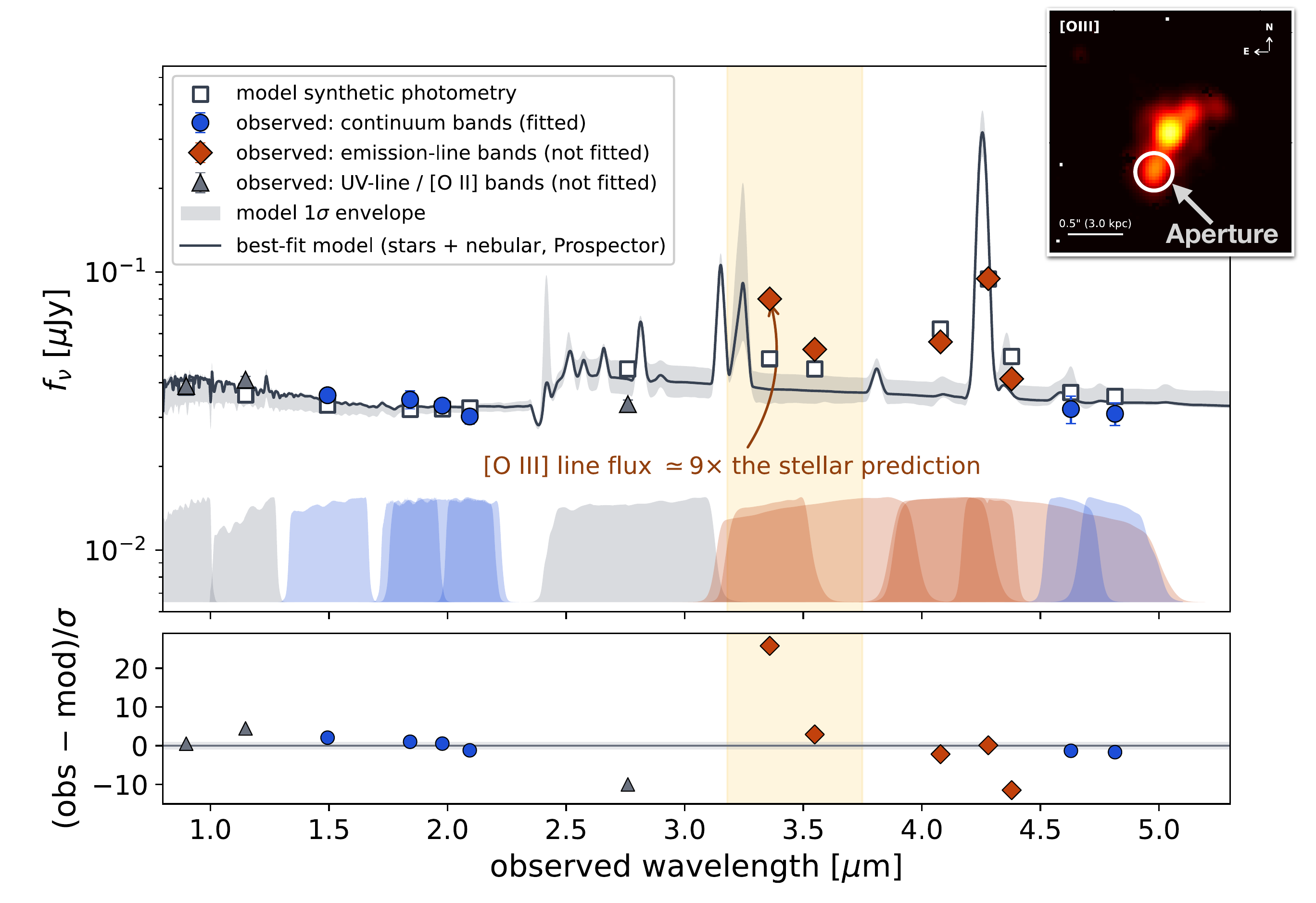}
    \caption{Continuum SED fit to the southeastern clump. Top: the dark curve and grey band show the best-fit \texttt{Prospector} model and the 1-$\sigma$ posterior range. Only the six line-free bands (blue circles) are used in the fit; the remaining bands contain strong emission lines and are shown for comparison. Open squares show the corresponding model photometry. Bottom: residuals between the observed and model photometry. F335M, which contains \mbox{[\ion{O}{3}]}, shows a strong excess, corresponding to an observed \mbox{[\ion{O}{3}]} flux approximately nine times higher than the median stellar-photoionization prediction.}
    \label{fig:cont_sed}
\end{figure}

As shown in Figure \ref{fig:cont_sed}, a stellar-population model can reproduce the continuum measurements. We then use the posterior models to predict the line emission associated with stellar photoionization. The predicted Balmer emission is also consistent with the observed fluxes, whereas the observed \mbox{[\ion{O}{3}]} flux is significantly higher than the model prediction. Thus, if the observed continuum is associated with a stellar population in the southeastern clump, its stellar photoionization can account for the Balmer emission but not for the strong \mbox{[\ion{O}{3}]} emission. An additional high-excitation ionizing contribution is therefore required.

In summary, our NIRCam optical line maps reveal a compact, PSF-consistent H$\alpha$+\mbox{[\ion{N}{2}]} core at the LRD
continuum centroid, an extended component along a southeast-to-northwest axis traced by these optical lines and
matching the elongated UV continuum, and a bright \mbox{[\ion{O}{3}]}-dominated clump at the southeastern end of this axis where the $\mbox{[\ion{O}{3}]} / (\text{H}\alpha + \mbox{[\ion{N}{2}]})$ ratio peaks.  This same southeastern region is aligned with the MUSE blue Ly$\alpha$ and tentative \mbox{N\,{\sc v}} emission. Continuum SED modeling further shows that, while a stellar-population model can reproduce the continuum and Balmer emission of the south-eastern clump, it underpredicts its \mbox{[\ion{O}{3}]} emission significantly. We will elaborate on the physical origin of these features in Section~\ref{sec:discussion}.

\section{Discussion and Summary}
\label{sec:discussion}

The central engine of LRD-204851 has already been characterized in detail elsewhere, as reviewed in Section~\ref{sec:intro}. The combined picture is broadly consistent with
recent models invoking dense neutral gas clumps around AGNs or ``black-hole-star''-like configurations for the broad-line emission of LRDs \citep[e.g.,][]{Inayoshi2025, Naidu2025,cliff}, in which the central black hole is enshrouded by a dense, partially ionized gas envelope of substantial column density. In LRD-204851 itself, spectroscopic evidence for slow outward motion of such a central envelope, of order
100\,km\,s$^{-1}$, is provided by the P-Cygni Balmer absorber identified in a very recent NIRSpec R2700 spectrum by \citet{Matthee2026}.
Adopting this picture, we focus here on the question that the present study uniquely constrains: the origin of the remarkable
$\sim$kpc-scale elongated structure of LRD-204851, made visible in detail
by the combination of MUSE rest-frame UV spectroscopy and JWST/NIRCam optical emission-line mapping.

Our combined MUSE and NIRCam data consistently support a scenario in which the central engine of LRD-204851 illuminates a
$\sim$kpc-scale region of its host galaxy through low-$N_{\rm HI}$ channels with small opening angles. The clearest empirical signature of this geometry is the thin, bipolar, elongated structure visible in the NIRCam line maps and rest-frame UV continuum, which passes through the F460M continuum centroid and extends several kpc on either side of it along a southeast-to-northwest axis (Figure \ref{fig:target_info} and Section~\ref{sec:nircam}). Along the same axis, three additional empirical diagnostics align: the spatial offset of the blue Ly$\alpha$ component, the location of the tentative \mbox{N\,{\sc v}} peak, and a bright [\mbox{O\,{\sc iii}}] clump-like structure at the south-eastern end of the elongated structure and spatially aligned with the tentative \mbox{N\,{\sc v}} emission.  Together these features are naturally accommodated by the favored \texttt{Bicone\_X\_Slab\_Out} \texttt{zELDA} Ly$\alpha$ modeling results. The physical origin of this bipolar elongation, however, is not yet uniquely determined.  Below we consider three configurations that
are each consistent with the observed geometry.

The channel may have been actively carved by an outflow launched
from the central engine, with the south-eastern clump tracing the
working surface where outflowing material encounters the ambient
ISM and is shock- or photo-ionized to the high values of
[\mbox{O\,{\sc iii}}]/\mbox{H\,$\alpha$} we observe.  Empirically, the very thin, bipolar morphology
of the elongated structure requires a high
degree of geometric collimation.  This collimation can arise either
intrinsically, as in a jet, with the resulting kpc-scale
ionization structure morphologically reminiscent of those observed
in low-redshift Seyferts and radio galaxies
\citep[e.g.,][]{McCarthy1993,Carilli1996,Tadhunter2016}, or
geometrically, in which a wider-angle wind from the central engine
is funneled through the narrow low-$N_{\rm HI}$ biconical opening of
the surrounding dense cocoon. The current data do not discriminate
between these two routes to collimation.

The kinematic constraints on this outflow remain modest.  Our Ly$\alpha$ modeling yields a bulk expansion velocity $V_{\rm exp}\!\simeq\!110$\,km\,s$^{-1}$ in the favored biconical geometry (Section~\ref{sec:lya_rt}), in agreement with the P-Cygni \mbox{H\,$\alpha$} absorber at $\Delta v \approx -80$\,km\,s$^{-1}$ \citep{Matthee2026}.  We
emphasize that this outflow is geometrically anisotropic and biconical, distinct from the spherically symmetric \texttt{Galactic\_Wind} configuration that our \texttt{zELDA} model
comparison decisively disfavors (Appendix~\ref{app:geometries}); the disfavored \texttt{Galactic\_Wind} model reflects a geometric mismatch with the
observed two-peak Ly$\alpha$ profile rather than the absence of outflow.  Combining the intrinsic Ly$\alpha$ emission velocity offset
($\Delta v \approx -130$\,km\,s$^{-1}$ relative to systemic) with $V_{\rm exp}$, the bulk near-side cone gas reaches a systemic-frame velocity of approximately $-240$\,km\,s$^{-1}$, which sits at the
lower edge of the $v_{\rm s}\!\gtrsim\!200$--$300$\,km\,s$^{-1}$ threshold required for collisional shock ionization of high-ionization species such as \mbox{N\,{\sc v}}
\citep[e.g.,][]{Dopita1996, Allen2008,
Sutherland2017}.  Both photoionization by hard radiation escaping the central engine along the low-$N_{\rm HI}$ channels and
modest shock ionization in the bulk outflow are therefore plausible mechanisms for the high-ionization line emission at the
south-eastern clump.  If the tentative
\mbox{N\,{\sc v}}\,$\lambda 1238$ detection is confirmed at higher S/N, its central velocity ($\Delta v \approx -467$\,km\,s$^{-1}$ relative to systemic) would lie above the shock-ionization
threshold, and both production channels would remain viable.  We further note that, if confirmed, this \mbox{N\,{\sc v}} velocity
exceeds the bulk cone-gas value implied by combining $z_{\rm Ly\alpha}$ and $V_{\rm exp}$.  This difference can be accommodated by the resonant nature of Ly$\alpha$, whose observed peak positions are a non-trivial function of $V_{\rm exp}$,
$N_{\rm HI}$, geometry, and the intrinsic line profile, and can be reproduced by a range of bulk velocities once these parameters are varied
\citep[e.g.,][]{Verhamme2006, Verhamme2008}. 

It is worth noting that LRDs as a population have been reported to be radio-quiet at observed GHz frequencies \citep[e.g.,][]{Akins2025, Perger2025}. LRD-204851 itself is undetected in the ultra-deep VLA 3 and 6\,GHz
observations of the GOODS-S/HUDF region presented by \citet{Lyu2022} and \citet{Alberts2020}, which reach pointing-center rms sensitivities of $0.75$ and $0.32\,\mu$Jy beam$^{-1}$, respectively, and remain among the deepest available at these frequencies.  At $z=5.482$, however, these observations probe rest-frame frequencies of
$\approx 20$ and $40$\,GHz.  Non-thermal jet emission typically has a non-flat radio spectrum, either rising at low frequencies in
synchrotron-self-absorbed compact young sources \citep[e.g.,][]{ODea1998} or steeply falling at high frequencies in optically thin extended lobes \citep[e.g.,][]{Begelman1984}; in both regimes the brightest non-thermal emission lies below the rest-frame frequencies probed here. Deep, lower-frequency radio observations would therefore be required to firmly constrain a low-power non-thermal jet contribution to the structure observed here.

A second possibility is that the channel is a quasi-static cavity in a clumpy host ISM, illuminated by hard photons escaping the central envelope. The south-eastern clump would then represent gas photoionized by the present-day or recently faded ionizing radiation of the central engine, akin to extended narrow-line regions in
low-redshift Type-2 AGN \citep[e.g.,][]{Unger1987, Mulchaey1996,
StorchiBergmann2018}, similar structures recently discovered at $z > 5$ with JWST/NIRSpec IFU
\citep[e.g.,][]{Wylezalek2022, Cresci2023, Ubler2024, Lyu2025}, and ``ionization echo'' systems such as Hanny's Voorwerp
\citep{Lintott2009}.  The thin, bipolar morphology is naturally produced by a narrow cone opening angle.  The static-cone
interpretation, however, predicts that the south-eastern gas itself sits near the host systemic velocity, and would be directly disfavored by either a confirmed blueshifted \mbox{N\,{\sc v}} emission or spatially resolved [\mbox{O\,{\sc iii}}] kinematics along the elongated structure revealing significant velocity gradients.

A third possibility is that the elongated structure is a pre-existing host-galaxy feature, such as a tidal arm, an in-situ star-forming filament, or a low-mass companion, that happens to lie along the cone axis. The continuum SED of the southeastern clump can indeed be reproduced by a stellar-population model (Section \ref{sec:cont_sed}), so the presence of such a host-galaxy component cannot be excluded. However, the same model reproduces the Balmer emission while underpredicting [\mbox{O\,{\sc iii}}] significantly, indicating that in-situ star formation alone cannot explain the ionization state of the structure. An additional high-excitation source, such as radiation from the central engine or shocks, is still required. Moreover, the host feature would have to be aligned with the cone axis and with the blue Ly$\alpha$, tentative \mbox{N\,{\sc v}}, and high-ionization [\mbox{O\,{\sc iii}}] emission. We therefore consider a purely star-forming host feature to be unlikely, although a pre-existing stellar structure affected by the central engine cannot be excluded.

To summarize, we have combined MUSE rest-frame UV integral-field spectroscopy with NIRCam rest-frame optical emission-line maps
to characterize the gas geometry, kinematics, and ionization state of LRD-204851 at sub-kpc resolution. The MUSE Ly$\alpha$ profile is
best described by Ly$\alpha$ radiative transfer through an expanding, dense neutral envelope pierced by a biconical low-column-density cavity, and the same data reveal a tentative blueshifted \mbox{N\,{\sc v}}\,$\lambda 1238$ detection.  The NIRCam
line maps further reveal a thin, bipolar, elongated structure passing through the optical continuum centroid and extending several kpc on either side along a southeast-to-northwest axis, with a bright high-ionization [\mbox{O\,{\sc iii}}] clump-like structure at the south-eastern end aligned with the cone-leakage axis inferred from the Ly$\alpha$ modeling.  While the current data do not yet uniquely distinguish among the possible physical origins of this elongated structure, the analyses presented here converge on a common picture in which the elongated emission of LRD-204851 is connected to radiation and/or gas flow from its central engine through a narrow, low-column-density biconical channel. LRD-204851 therefore represents one of the clearest currently known cases in which the impact of an LRD central engine on its host galaxy may be directly observable on $\sim$kpc scales.

\section*{Acknowledgments}

ZJ, YS, YZ, GHR, JL, MR and ST acknowledge support from the NIRCam Science Team contract to the University of Arizona, NAS5-02015. The work of CCW is supported by NOIRLab, which is managed by the Association of Universities for Research in Astronomy (AURA) under a cooperative agreement with the National Science Foundation.

\appendix 

\section{Astrometric correction for the MUSE data}\label{app:astrometry}

The astrometric solution of the MUSE-Wide mosaic is anchored to the Gaia-tied JADES NIRCam frame using a source-based registration to the JADES F090W mosaic. Reference sources are selected from the JADES DR5 photometric catalog \citep{Robertson2026} within the MUSE field of view, requiring them to be bright ($18 < m_{\rm F606W} < 24$). Sources separated from any other catalog neighbor by less than $0.8''$, comparable to the typical seeing of the MUSE HUDF observations \citep{Bacon2023}, are excluded to avoid blends in the MUSE data. For each of the $N$ retained sources, we compute centroids in both a white-light collapse of the MUSE cube and the corresponding NIRCam F090W cutout using \texttt{photutils.centroids.centroid\_com}. After requiring a centroid peak S/N $>3$ in the MUSE white-light image, we retain $N\approx1200$ sources for the astrometric correction.

\begin{figure}
    \centering
    \includegraphics[width=0.77\linewidth]{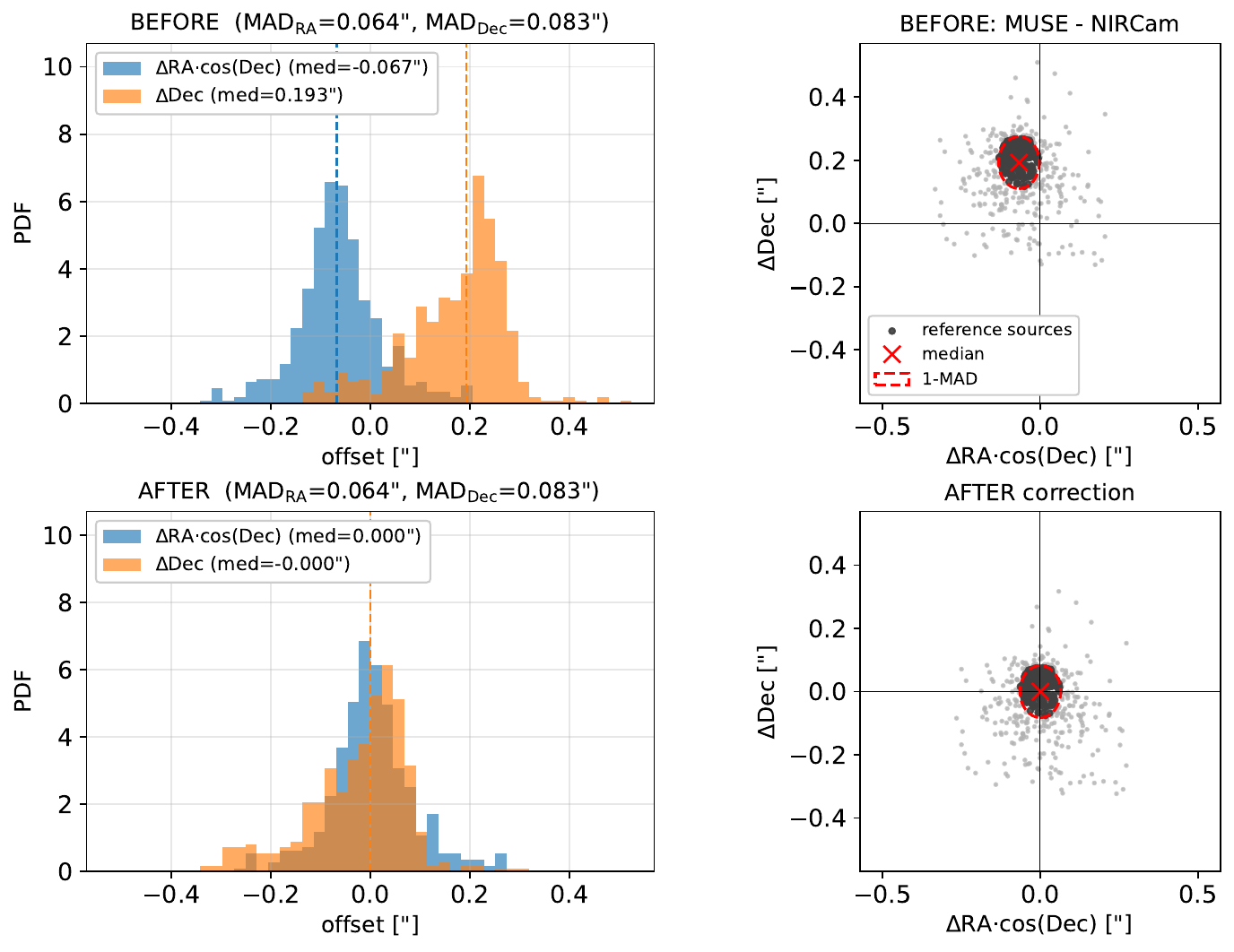}
    \caption{MUSE-to-NIRCam astrometric correction.  \textbf{Left}: PDF of the per-source offsets for reference sources before (top) and after (bottom) correction.  \textbf{Right}: 2-D astrometric residuals; the red dashed ellipse marks $\pm$1\,mean absolute deviation (MAD) about the median. After correction both medians are zero.}
    \label{fig:wcs}
\end{figure}

The per-source sky offsets are sigma-clipped at $3\sigma$, and their median gives a global astrometric correction of
$(\Delta\alpha\cos\delta,\,\Delta\delta)=(-0.0675'',\,+0.1926'')$.
A four-parameter quality-assurance fit, allowing for shift, rotation, and scale, confirms that the rotation and scale residuals are negligible
($\theta < 0\fdg005$, $\Delta s < 3\times10^{-5}$); we therefore apply only a pure shift. Figure~\ref{fig:wcs} shows the improvement after this astrometric correction: the median relative offset between JADES NIRCam and MUSE is consistent with zero, with a mean absolute deviation of $\sim0.07''$.

\section{Comparison of the four \texttt{zELDA} geometries}
\label{app:geometries}

\begin{figure}
\centering
    \includegraphics[width=0.47\textwidth]{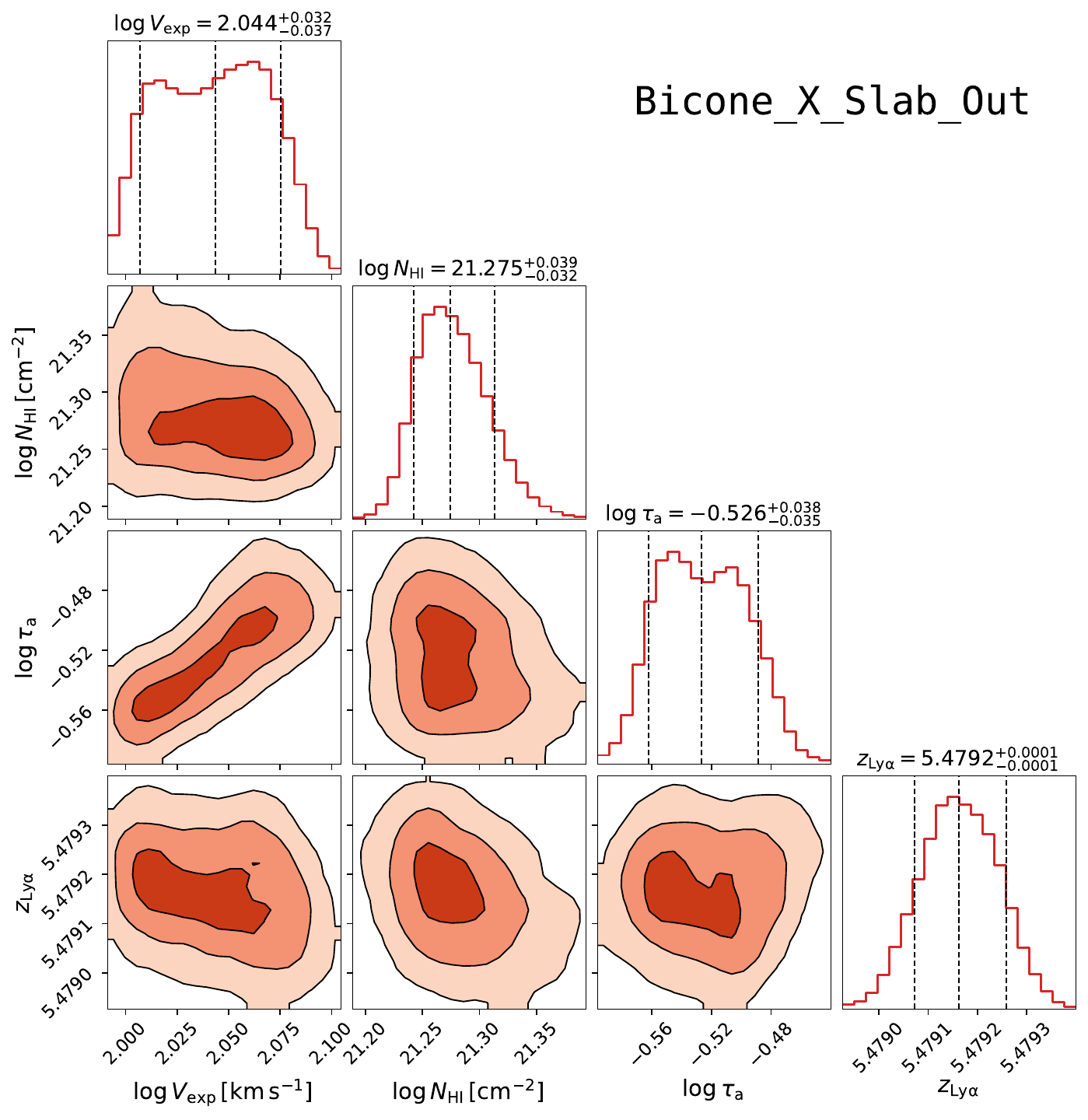}
    \caption{Joint and marginal posterior distributions for the four sampled parameters, $z_{\rm Ly\alpha}$, $\log V_{\rm exp}$, $\log N_{\rm HI}$, and $\log\tau_{\rm a}$, under the \texttt{Bicone\_X\_Slab\_Out} geometry, which best reproduces the observed Ly$\alpha$ profile of LRD-204851.}
    \label{fig:zelda_corner}
\end{figure}

In Section~\ref{sec:lya_rt}, we adopt the \texttt{Bicone\_X\_Slab\_Out} geometry as our fiducial model; its full posterior is shown in Figure~\ref{fig:zelda_corner}. Here we report the corresponding fits under the three alternative \texttt{zELDA} geometries, \texttt{Thin\_Shell}, \texttt{Galactic\_Wind}, and \texttt{Bicone\_X\_Slab\_In}, and quantify their relative quality. All four runs use identical fitting windows ($7855$--$7950\,\AA$ in the observed frame), identical priors, and identical nested-sampling settings in \texttt{dynesty}. The priors are flat in $\log V_{\rm exp}\,[{\rm km\,s^{-1}}]\in[1.0,3.0]$, flat in $\log N_{\rm HI}\,[{\rm cm^{-2}}]\in[17,22]$, flat in $\log\tau_{\rm a}\in[-3,1]$, and Gaussian in $z_{\rm Ly\alpha}$, centered at $z_{\rm sys}=5.482$ with $\sigma_z=0.05$. Because the four geometries share the same parameter dimensionality and priors, model comparison reduces to a straightforward Bayes-factor calculation using the \texttt{dynesty} marginal log-evidence, $\log\mathcal{Z}$ \citep{Kass1995}.

Relative to the fiducial \texttt{Bicone\_X\_Slab\_Out} run, we find
$\Delta\log\mathcal{Z}=-8.2$ for \texttt{Galactic\_Wind},
$\Delta\log\mathcal{Z}=-16.3$ for \texttt{Thin\_Shell}, and
$\Delta\log\mathcal{Z}=-94.6$ for \texttt{Bicone\_X\_Slab\_In}. All three
alternatives are therefore significantly disfavored by the data
($|\Delta\log\mathcal{Z}|\!>\!5$ corresponds to
``decisive'' evidence, and
$|\Delta\log\mathcal{Z}|\!>\!10$ to overwhelming evidence according to \citealt{Kass1995}). 

\begin{figure*}
\centering
    \includegraphics[width=\textwidth]{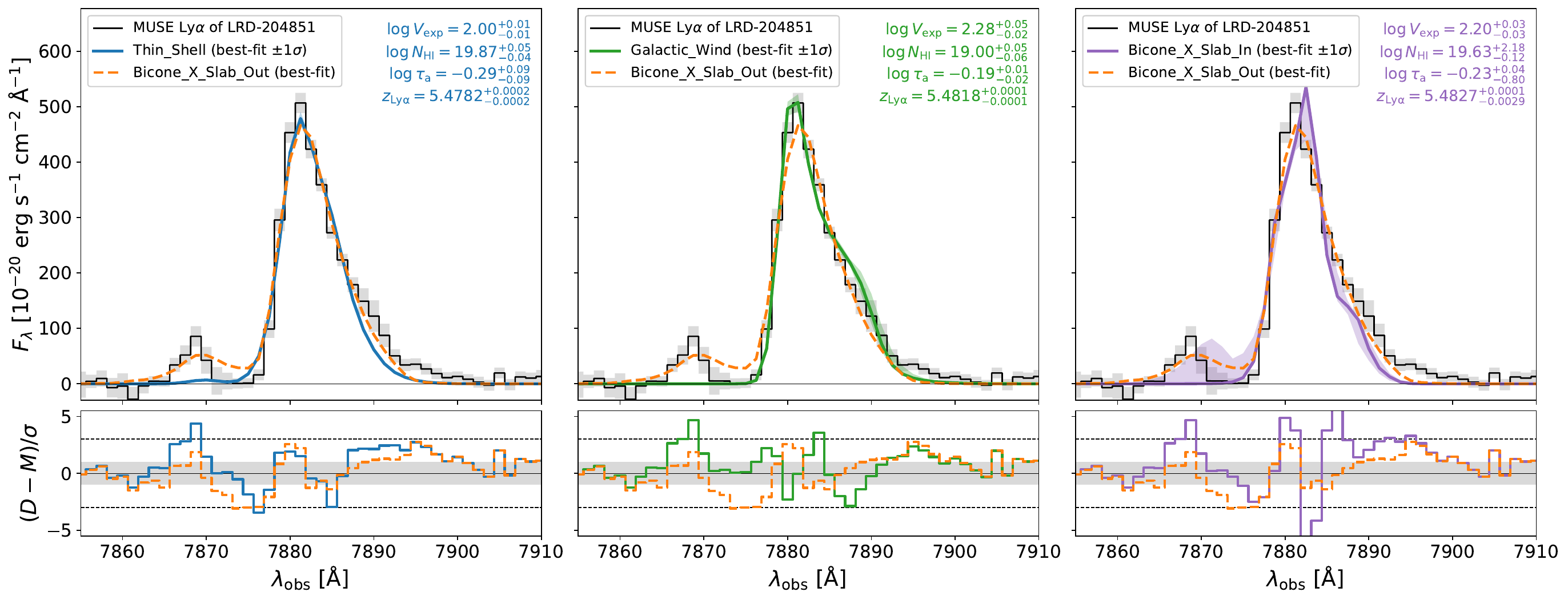}
    \caption{Best-fit \texttt{zELDA} Ly$\alpha$ profiles for the three alternative gas geometries, overlaid on the observed MUSE Ly$\alpha$ spectrum (black, with $1\sigma$ shading). The \texttt{Bicone\_X\_Slab\_Out} model is shown in all panels as a red dashed line for comparison. The bottom panels show the residuals, demonstrating that \texttt{Bicone\_X\_Slab\_Out} outperforms the other three geometries.}
    \label{fig:zelda_3geom_fits}
\end{figure*}

The driver of these differences is apparent in Figure~\ref{fig:zelda_3geom_fits}. Both spherically symmetric
geometries (\texttt{Thin\_Shell} and \texttt{Galactic\_Wind}) reproduce
the dominant red Ly$\alpha$ peak only by adopting a low neutral-hydrogen
column ($\log (N_{\rm HI}/\rm cm^{-2})\!\simeq\!19$); they fail to produce a
sufficiently blueshifted, narrow companion peak, because back-scattering through a
uniform shell or wind always places the secondary peak much closer to
systemic. The \texttt{Bicone\_X\_Slab\_In} geometry (sight-line down the
cone axis, i.e. face-on) suppresses the resonant-scattering pathway that would build
the strong red peak, and its posterior runs to low $N_{\rm HI}$ and a
large $\tau_{\rm a}$ in an unsuccessful attempt to suppress the
unobserved blue wing. Only \texttt{Bicone\_X\_Slab\_Out} offers a
two-channel escape: a high-$N_{\rm HI}$ static slab through which
Ly$\alpha$ photons back-scatter to form the red peak, and a low-column
biconical cavity that allows a small fraction of photons to escape with
a much larger blueshift. This configuration naturally reproduces both
the $\sim\!400\,{\rm km\,s^{-1}}$ separation between the two observed
peaks and their $\sim\!6\%$ flux ratio.

\bibliography{sample701}{}
\bibliographystyle{aasjournalv7}



\end{document}